\documentclass[11pt,twoside]{article}
\usepackage{asp2010}
\def\gapprox{\;\rlap{\lower 2.5pt            
 \hbox{$\sim$}}\raise 1.5pt\hbox{$>$}\;}
\def\lapprox{\;\rlap{\lower 2.5pt            
 \hbox{$\sim$}}\raise 1.5pt\hbox{$<$}\;}
\def\N{\,{\rm I\kern-.20em N}}

\resetcounters

\begin{document}

\title{The Radio--X-ray Relation in Cool Stars: Are we headed toward a divorce?}

\author{Jan Forbrich$^1$, Scott J. Wolk$^1$, Manuel G\"udel$^2$, Arnold Benz$^3$, Rachel Osten$^4$, Jeffrey L. Linsky$^5$, Margaret McLean$^1$, Laurent Loinard$^6$, and Edo Berger$^1$}
\affil{$^1$Harvard-Smithsonian CfA, 60 Garden St, Cambridge, MA 02138, USA}
\affil{$^2$University of Vienna, Department of Astronomy, T\"urkenschanzstr. 17, A-1180 Vienna, Austria}
\affil{$^3$Institute of Astronomy, ETH Zurich, 8093 Zurich, Switzerland}
\affil{$^4$Space Telescope Science Institute, 3700 San Martin Drive, Baltimore, MD 21218, USA}
\affil{$^5$JILA, University of Colorado and NIST, Boulder, CO 80309-0440, USA}
\affil{$^6$Centro de Radioastronom\'ia y Astrof\'isica, Universidad Nacional Aut\'onoma de M\'exico, Apartado Postal 3-72, 58090, Morelia, Michoac\'an, Mexico}

\begin{abstract}
This splinter session was devoted to reviewing our current knowledge of correlated X-ray and radio emission from cool stars in order to prepare for new large radio observatories such as the EVLA. A key interest was to discuss why the X-ray and radio luminosities of some cool stars are in clear breach of a correlation that holds for other active stars, the so-called G\"udel-Benz relation. This article summarizes the contributions whereas the actual presentations can be accessed on the splinter website\footnote{http://cxc.harvard.edu/cs16xrayradio/}.
\end{abstract}

\vspace*{-9mm}
\section{Radio emission and X-rays as diagnostics of coronal energy release}

\begin{figure}[t!]
\includegraphics[angle=0,width=9cm]{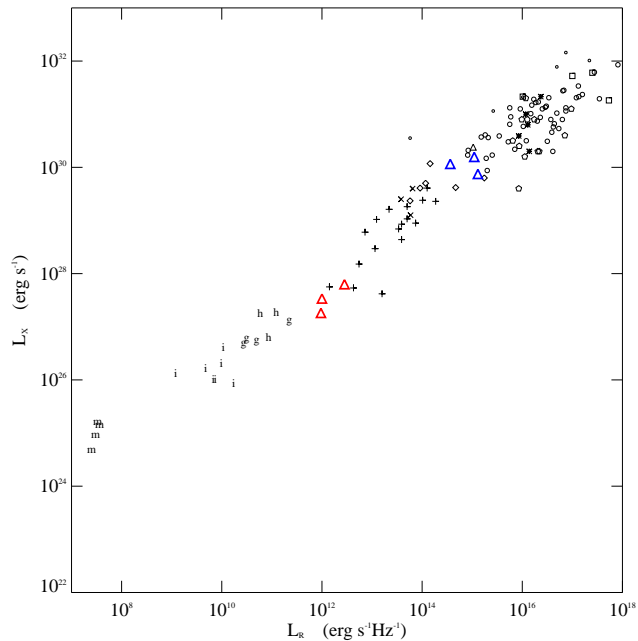}
\caption{Radio vs. X-ray correlation. Different symbols in the upper right part of the figure refer to
different types of active stars. The letters in the lower left show the loci of solar flares (luminosities
averaged over the duration of the X-ray flare; see \citealt{benz94} for details). The red and blue triangles 
show examples of stellar flares from M dwarfs \citep{guedel96} and RS CVn binaries \citep{osten04}.}
\label{fig1} 
\end{figure}

X-rays and radio emission are excellent diagnostic probes to study energy release in magnetized
stellar coronae. Solar observations have been key to deciphering the plethora of phenomena seen
in these wavelength ranges. In brief, X-rays trace the presence of dense, hot (million-degree) plasma
trapped in closed coronal magnetic fields, heated by processes that are still not fully understood.
Radio observations, in contrast, probe both thermal atmospheric components from the chromosphere to
the corona, and populations of non-thermal, accelerated electrons, typically
residing in low-density, open or closed coronal magnetic fields. 

A solar-stellar analogy is, however, complicated by phenomenology in {\it magnetically active stars} 
that is, at first sight, not  present in the Sun. X-ray emission becomes much stronger toward more active
stars, most likely as a result of increased surface coverage with active regions; however,  
the average, ``characteristic'' temperature of the corona increases along with the coronal luminosity
(e.g., \citealt{schrijver84, guedel97}), a trend that requires additional physical explanation. 
At radio wavelengths, magnetically active stars also show a different face. Solar
radio emission in the 1-10~GHz range is dominated by bremsstrahlung from various
chromospheric/transition region levels and by optically thick gyroresonance emission from coronal layers above 
magnetic active regions. In contrast, observed radio brightness temperatures and radio spectra from active 
stars indicate gyrosynchrotron radiation from electrons with much higher energies than typically present
in the solar atmosphere. The Sun occasionally features gyrosynchrotron emission, often accompanied
by a variety of coherent radiation types (see Sect.~\ref{sect_benz}), but such radiation is confined to 
episodes of flaring.

Magnetically active stars reveal further radio properties lacking any clear solar analogy such as
extremely large coronal structures with size scales of order a stellar radius and more
(e.g., \citealt{benz98, mutel98, peterson10}). In contrast, X-ray coronae tend to be rather compact even in 
extremely active stars (e.g., \citealt{walter83, ottmann93}); this is a consequence of the X-ray brightness 
scaling with the square of the electron density combined with relatively small pressure scale heights. 
Radio and X-ray sources are therefore not necessarily co-spatial
and probe rather different plasmas or particle populations
in stellar atmospheres, different atmospheric layers and structures, and perhaps even different energy
sources. There should be little reason to expect that the two types of radiation are correlated
in stars. 

It therefore came as a surprise when such a correlation was uncovered for the steady radio and
X-ray luminosities of RS CVn close binary stars. \citet{drake89} found that the 
soft X-ray and radio (6~cm) luminosities are correlated over a few
orders of magnitude albeit somewhat different from a linear trend ($L_{\rm R} \propto L_{\rm X}^{1.37\pm 0.13}$).
They suggested a self-consistent 
scheme in which the radio emission originates from the tail of the Maxwellian electron distribution of a 
very hot ($>$50~MK) plasma through the gyrosynchrotron process; such a plasma component is 
suggested from X-ray observations. This model is elegant as it links X-rays and radio emission by
suggesting a common source. However, serious difficulties remain.
Gyrosynchrotron spectra from a thermal plasma reveal a steep decline toward higher frequencies not observed 
in any magnetically active star; an acceptable spectral fit requires an extraordinary setup of the
coronal magnetic field, such as a field strength decreasing with radius as $r^{-1}$ \citep{chiuderi93, beasley00}.  
Non-thermal (power-law) electron distributions, in contrast, readily produce shallow spectra as observed,
especially if the ``aging'' of an injected electron distribution, leading to spectral modifications due to
synchrotron and collisional losses, is taken into account \citep{chiuderi93}.

A closer inspection of stars more akin to the Sun is in order. To that end, \citet{guedel93a} studied
X-ray and radio luminosities for M dwarfs, followed by other spectral types including G dwarfs (e.g., 
\citealt{benz94, guedel95}). Again, active stars of all late-type spectral classes followed a similar trend, 
best described by a proportionality, $L_{\rm X}/L_{\rm R} \approx 10^{15.5\pm 0.5}$~Hz (in the following referred to as the GB relation). Combining
these samples with samples of RS CVn binaries, Algol binaries, FK Com-type stars and also pre-main 
sequence weak-lined T Tauri stars, a coherent trend is found over 5-6 orders of magnitude in $L_{\rm R}$ and
$L_{\rm X}$ (Figure~\ref{fig1}). It is important to note that the $L_{\rm X}/L_{\rm R}$ ratio is by no 
means  universal. It has been demonstrated
exclusively for {\it magnetically active stars} but does clearly not apply to inactive stars like the Sun;
such stars keep appreciable levels of quasi-steady soft X-ray emission but are {\it not} sources of continuous
radio emission of the gyrosynchrotron type. In fact, present-day radio observatories still cannot systematically
detect nearby cool stars except for extremely active examples.

Active stars stand out by two properties mentioned above - their 
extremely hot plasma seen in X-rays, and their non-thermal electron populations evidenced by their radio emission.
Let us assume that the  energy initially contained in the accelerated electrons eventually heats the coronal 
plasma. If the corona releases energy at a rate $\dot{E}$ by injecting accelerated electrons at an energy-dependent 
rate $\dot{n}_{\rm in}(\epsilon)$, then 
\begin{equation}\label{equil}
\dot{E} = {1\over a}\int_{\epsilon_0}^{\infty} \dot{n}_{\rm in}(\epsilon)\epsilon d\epsilon = {1\over b}L_{\rm X} 
\end{equation}
where $a$ is the fraction of the total energy that is channeled into particle acceleration and $b$ is the
fraction of the total energy that is eventually radiated as soft X-rays. Eq.~\ref{equil} assumes an equilibrium
between energy injection and energy loss. After introducing radiation processes into Eq.~\ref{equil}, one finds
\citep{guedel93b}
\begin{equation}\label{correl} 
L_{\rm R} = 3\times 10^{-22}B^{2.48}{a\over b}\tau_0 (\alpha + 1) L_{\rm X}
\label{luminosities}
\end{equation}
i.e., a proportionality if several parameters on the right-hand side take characteristic, constant values,
in particular $B$ and the ratio $a/b$ ($\alpha$ is the power-law index for the energy dependence of the
electron lifetime). Conversely, comparing Eq.~\ref{correl} with observations, one
finds (e.g., for $a/b\approx 1$) the time scale $\tau_0$ for electron trapping (i.e., the lifetime of the
population), concluding that the radiation must decay on time scales of minutes to hours in most cases.
This necessitates frequent or quasi-continuous replenishment of the corona by high-energy
electrons.

This mechanism is what the ``standard model'' for a solar flare would predict. The standard solar flare
model, the chromospheric evaporation scenario, posits that electrons initially accelerated in reconnecting 
magnetic fields propagate to chromospheric layers where they heat and ablate material which escapes 
into closed magnetic loops and cools by X-ray radiation. The best observational evidence for this
model is the ``Neupert Effect'', stating that the time derivative of the flare X-ray
light curve resembles the radio (or hard X-ray or U-band) light curve, $dL_{\rm X}/dt\propto L_{\rm R}$.

This prediction follows from assuming that $L_{\rm X}$ roughly scales with the thermal energy content in the 
hot plasma accumulated from the high-energy electrons, while radio emission scales with the number of such 
electrons present at any given time. The Neupert Effect is frequently observed in solar flares 
(e.g., \citealt{dennis93}), but has also frequently been seen in stellar flares, both 
extremely large events and the smallest yet discerned in stellar X-rays  (e.g., 
\citealt{guedel96, guedel02, osten04}).

We need one further ingredient, relating
flares to the observed quiescent emission. During the past decade, a number of studies have shown that
the occurrence rate of stellar flares in X-rays is distributed as a power law in radiated energy (a concept
familiar to solar physics), $dN/dE \propto E^{-\alpha}$,
with $\alpha \ga 2$ (e.g., \citealt{audard00, kashyap02}). In that case, and assuming that the
power law continues toward smaller energies, the energy integration, $ \int_{\rm E_0}^{\infty}E(dN/dE)dE$
diverges for $E_0\rightarrow 0$, i.e., a lower cut-off is required. More relevant here, the entire
apparently steady emission level could be explained by the large number of small flares that superpose
to a quasi-steady emission level while not recognizable individually in light curves. 

Assembling the above pieces, we then suggest to solve the $L_{\rm X}-L_{\rm R}$ puzzle as follows:
Radio and X-ray emission correlate in magnetically active stars because the radiation we perceive
as ``quiescent'' emission is made up of contributions from numerous small flares; each of these flares
heats plasma by transforming kinetic energy from accelerated electrons; a portion of the latter is evident from
their radio emission, while the heated plasma is observed by its X-ray emission. The $L_{\rm X}/L_{\rm R}$
ratio therefore reflects the energy loss ratio of individual flares.
As a check, we consider whether {\it solar and stellar flares} reveal radiative output ratios similar to those 
of the  ``quiescent'' radiation. Average X-ray and radio luminosities for a range of solar flares (with specified 
total flare durations) as well as a sample of stellar flares are overplotted in Fig.~\ref{fig1}. Indeed, the solar flares continue the trend seen in magnetically active stars \citep{benz94} and
the stellar flares show luminosity ratios in perfect agreement with the trend for quiescent emission.
These observations support a picture in which flares are at the origin of coronal heating, of the steady 
radiation in magnetically active stars, and consequently of the $L_{\rm X}/L_{\rm R}$ correlation.

\vspace*{-5mm}
\section{The Sun}
\label{sect_benz}
\vspace*{-1mm}
\paragraph{Correlation of Solar and Stellar X-rays with Gyrosynchrotron Radio Emission}

X-ray emission from stellar and solar flares and coronae is produced by bound-bound transitions and by bremsstrahlung of rapidly moving electrons deflected on ions. The emission comes in two flavors depending on the energy distribution of the electrons: thermal or non-thermal. Thermal X-rays range from less than 0.1 keV to more than 10 keV, resulting from temperatures in the range between $10^{6-8}$ K. Non-thermal X-rays are emitted by energetic electrons, accelerated by plasma processes. Their bremsstrahlung emission is observed in solar flares from about 10 keV up to 100 MeV. It is approximately a power law in the photon energy spectrum, often with some breaks caused by a change in power-law index. The observed spectrum indicates a power law also in the energy distribution of electrons.

Gyrosynchrotron radio emission is produced by individual particles, usually mildly relativistic ($>$ 100 keV) non-thermal electrons. The emission is caused by their spiraling motion in magnetic fields (e.g., \citealp{dulk82}).
Each high-energy electron radiates both bremsstrahlung and gyrosynchro\-tron emission. Although different parameters enter the emissivity (most notably the magnetic field in the gyrosynchrotron case), it is not surprising that non-thermal X-rays and gyrosynchrotron emission correlate in solar flares. \citet{kosugi88} find a linear correlation between non-thermal X-ray and radio peak fluxes with deviations of less than a half order of magnitude.
It is more surprising that in both
stellar quiescent and flaring coronae, the {\it thermal} (soft)
X-ray luminosity also correlates with gyrosynchrotron radio
emission (see above).

The correlation of non-thermal gyrosynchrotron radiation and thermal X-ray emission fits the standard flare scenario of flares: A significant fraction of the flare energy is released in the form of non-thermal electrons, which precipitate to a dense medium and heat it to the point of thermal X-ray emission. The scenario has not been confirmed in solar flares at a quantitative level.
It is surprising, nevertheless, that over the large range and widely different objects the correlation remains within half an order of magnitude \citep{krucker00}. In particular, the magnetic field and electron life time are expected to change. Some deviations are noted: RS CVn binaries, Algols, and BY Dra binaries tend to be radio-rich \citep{guedel93b}, nanoflares are radio-poor \citep{krucker00}. These differences may be the result of different parameters in Eq. (\ref{luminosities}).

The excellent correlation of the radio/X-ray relation has several consequences:

\noindent
1) The identical relation for flares and active star quiescent X-ray emission strongly suggests that their X-ray emitting corona is flare heated. 

\noindent
2) The radio/X-ray relation has allowed the discovery of radio emission of K, G, and F stars. Selecting bright X-ray emitters among these stellar types, they were easily detected for the first time in radio emission \citep{guedel94,guedel02araa}.

\noindent
3) A large deviation from the relation can be used to identify radio emission originating by a mechanism {\it other than gyrosynchrotron} (e.g., \citealp{benz01}).

\begin{figure}
\begin{center}
\leavevmode
\mbox{\hspace{-1cm}\epsfxsize=10.5cm
\epsffile{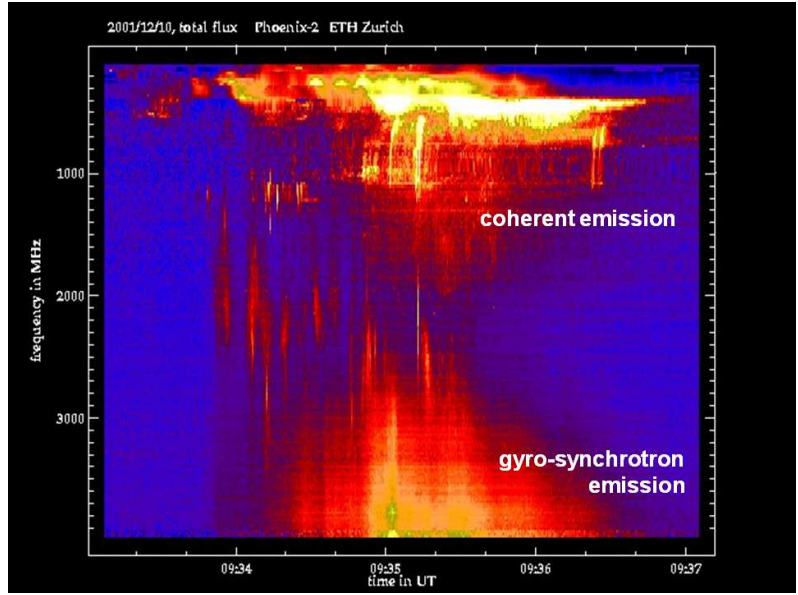}}
\end{center}
\caption[]{Spectrogram of solar flare radio emission in meter and decimeter waves observed with Phoenix-2 at ETH Zurich. The frequency increases downward. The example shows a clear separation of incoherent gyrosynchrotron emission at high frequencies and various coherent emissions at low frequencies.}
\label{fig_coherent}
\end{figure}

\vspace*{-3mm}
\paragraph{Correlation of Solar Soft X-rays with Coherent Radio Emission}
Cosmic radio emission originates in two ways: coherent and incoherent. Gyrosynchrotron and thermal radiation are incoherent and result from the emission of individual particles. Conversely, coherent emission is produced by a group of particles emitting in phase. The phasing may be caused by a wave in the plasma driven by an instability. Such instabilities occur in plasmas where electrons have non-thermal velocity distributions, such as a beam component, a loss-cone or  strong electric current \citep{benz02}. In wave terminology, the plasma wave is transformed into radio waves \citep{melrose80}. Excitation and transformation are highly non-linear. Thus the correlation with X-rays is weak or absent. Characteristics of coherent radio emission are a narrowband spectrum ($\Delta\nu/\nu \ll 1$),  high polarization ($\gapprox$40\% ), and extremely high brightness temperature ($\gapprox 10^{10}$ K). An example of various coherent emissions is shown in Fig \ref{fig_coherent}. At frequencies above 3000 MHz, gyrosynchrotron radiation appears as diffuse broadband structures increasing to higher frequencies beyond the instrumental limit at 4000 MHz.

Coherent radio emissions from the Sun have been reported at all wavelengths longer than 3 cm. At meter wavelengths, they are classified as the well-known Type I to V of radio bursts. At shorter wavelength, the metric types gradually disappear except Type III bursts which have been identified at up to 8.5 GHz. Other types, classified as DCIM, dominate at decimeter wavelengths. They consist of broadband pulsations, patches of continuum or narrowband spikes. Figure 5 in \citet{benz09} displays the highest fluxes reported. Gyrosynchrotron emission (marked therein as IV$\mu$) dominates up to 10 cm, followed by DCIM emission possibly directly associated with flare energy release, and Type II emission caused by coronal shock waves dominating above 100 cm. We note that a flux density of $10^6$ solar flux units (sfu) at 1 AU corresponds to 100 $\mu$Jy at 50 pc.

Solar coherent radio emissions were extensively compared to X-ray observations by the RHESSI and GOES satellites in a survey by \citet{benz05}. The survey finds that 17\% of the X-ray flares larger than GOES class C5 are not associated with coherent radio emission between 0.1 and 4 GHz. A later study, however, showed that all of the "radio-quiet" flares occurred either near the limb or showed emission below 0.1 GHz \citep{benz07}. Thus all flares appear to be associated by coherent radio emission.

Figure \ref{fig_coherent} shows that gyrosynchrotron emission may be associated with coherent radio emission but not correlate in detail. Correlation has been investigated not with gyrosynchrotron emission, but non-thermal X-rays which can be taken as a proxy. Some cases with good temporal correlation have been reported by \citet{dabrowski09}. However, recent imaging observations show that even correlating coherent radio emission does not originate from the location of coronal X-ray emission where electron acceleration is expected (Benz, Battaglia \& Vilmer, \textit{in prep.}). The current interpretation of coherent radio emission still associates some of the DCIM emissions with particle acceleration, but not necessarily with the major acceleration site. The new imaging observations therefore require a flare model that has a much larger volume than the coronal X-ray sources currently observable. 

Coherent radio emissions from solar flares do not correlate with thermal X-rays in detail except for the `big flare syndrome' (larger flares being generally more luminous at all wavelengths). \citet{benz06} find that the total energy emitted in decimeter radio waves scales on the average with the peak soft X-ray flux as $E_{\rm radio} = 1.92 \times 10^8 F^{1.42}_{\rm SXR}  {\rm [erg]}$
where $E_{\rm radio}$ is in units of $10^{20}$ erg and $F_{\rm SXR}$ is the GOES flux from 1.8 to 12.4 keV in units of W/m$^2$. The correlation coefficient is 0.75, with a 95\%\ significance range between 0.34 and 0.92. The scatter is more than an order of magnitude.

In most active stars and planets where the plasma frequency, $\omega_p$, exceeds the electron gyrofrequency $\Omega_e$, coherent radio emission is dominated by gyromagnetic maser emission at the first harmonic
$\nu = {\Omega_e/2\pi} = 2.80\times 10^6 B\ \ \ [{\rm Hz}]$.
Thus coherent radio emission, although not a measure for flare importance, may be used to measure the magnetic field strength in stellar and planetary atmospheres.

\vspace*{-6mm}
\section{RS CVn, Algol, and BY Dra Systems}

\begin{figure}[]
\includegraphics[angle=90,scale=0.44]{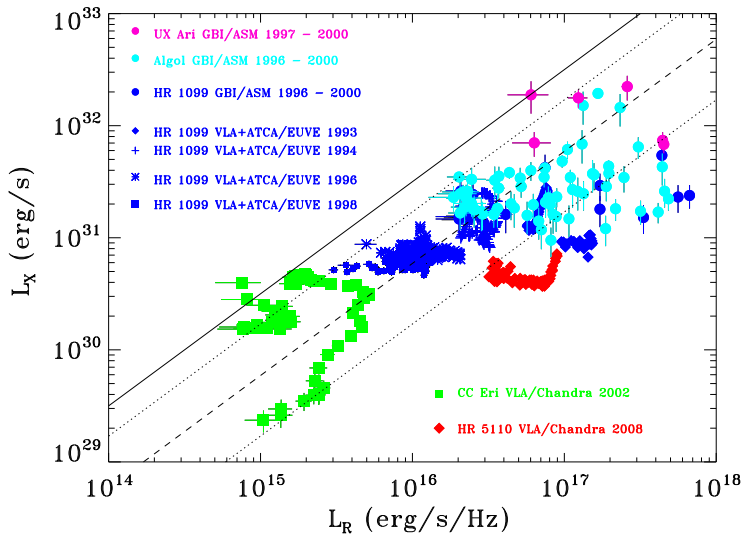}
\caption{X-ray vs. radio luminosity for different simultaneous multi-wavelength
observations of 5 different active stars: two Algol systems (Algol, HR 5110), 
two RS CVn binary systems (HR 1099 and UX Ari), and one BY Dra binary system
(CC Eri).  The dotted lines indicate the $L_{X}/L_{R}=\kappa\times10^{15.5\pm0.5}$ Hz ($\kappa$=0.17)
relationship in \citet{benz94}.  The dashed line gives the average
$L_{X}/L_{R}=$5.9$\times10^{14}$ Hz from these data, and the solid line
shows the GB relation with $\kappa$=1, $L_{X}/L_{R}=10^{15.5}$ Hz.}
\label{fig_ro}
\end{figure}

Active binaries consisting of RS CVn, BY Dra, and Algol systems
all lie at the upper right corner of the $L_X$ vs. $L_R$ diagram (Fig.~\ref{fig1}), exhibiting the extremes 
 of magnetic activity.  However,
these objects are also known to be highly time-variable sources.
This gives pause to whether relationships connecting time-averaged behaviors hold when strictly
simultaneous observations are used.

Based on simultaneous data summarized in the caption, Figure~\ref{fig_ro} shows the radio-X-ray correlation for five objects in comparison to the GB relation. Radio and X-ray variability are apparent, but most of the time
these hyperactive stars do display $L_{X}$ and $L_{R}$ values which are consistent with the
GB relation, albeit with a fair amount of scatter.  Note, however that the  range of variability differs between the two. For HR 1099 alone, $L_{R}$ spans
a range of 230 while only spanning a factor of 6 in $L_{X}$.

The conventional explanation for the nearly linear relationship between the two luminosities
argues that there is a common energy reservoir out of which both plasma heating and
particle acceleration occur, and these processes occur at a roughly fixed ratio between
different classes of active stars.  In the current situation, however, we can see that 
while this may be true in a time-averaged sense, the examples of uncorrelated radio and
X-ray flaring which produce the extreme values of $L_{X}/L_{R}$ have different ratios
of particle acceleration and plasma heating.  This is also true for flares 
which exhibit the Neupert effect (the green light curve of CC Eri contains one such
example), where the instantaneous $L{_X}/L_{R}$ ratio varies by a factor of
20 during the initial stages of the flare, spanning preflare conditions to 
the time when the flare-associated particle acceleration is 
occurring (radio luminosity peak) to the maximum flare-associated plasma heating (peak of the X-ray luminosity).

\vspace*{-6mm}
\section{Young Stellar Objects}

High-energy processes in Young Stellar Objects (YSOs) as observed in both the radio and the X-ray wavelength regime have been known for some time (e.g., \citealp{feigelson99}). Very early evolutionary stages of YSOs, class~I protostars, emit strong X-ray emission and nonthermal radio emission, presumably both due in large part to magnetically induced flares.
However, it is still unclear in which stage YSOs become radio-active. Since most YSOs are relatively weak radio continuum sources, an observational difficulty has been to unambiguously find genuine nonthermal radio sources, e.g., by their polarization. YSOs can also produce free-free thermal radio emission in ionized material, for example at the base of outflows or jets (for a review of YSOs in the context of stellar radio astronomy, see \citealp{guedel02}).

Over the last decade, YSO X-ray variability research has been put on a much improved statistical basis, particularly for short timescales. However, centimetric radio variability on similar scales has only been studied toward a few sources (e.g., \citealp{forbrich2008}). The correlation of variability in both wavelength ranges as well as the overall correlation of time-averaged radio and X-ray luminosities, as could be expected if the GB relation applies, remains unclear. While necessary to eliminate the effects of variability, there have only been a few simultaneous radio and X-ray observations.

\citet{feigelson94} obtained the first simultaneous radio and X-ray observations of a T Tauri star, targeting V773 Tau. They found uncorrelated variability, suggesting that the two emission mechanisms are decoupled. Following the discovery of the first class I protostar with nonthermal radio emission \citep{feigelson98}, the question of when protostars begin to be radio-active became more acute. \citet{gagne04} carried out a simultaneous radio and X-ray observing campaign of $\rho$ Oph, a region rich in YSOs of all classes. They detected several T Tauri stars in both wavelength ranges. The first simultaneous X-ray and radio detections of class~I protostars were reported by \citet{forbrichea2007} (see also \citealp{forbrichea2006}), targeting the \textit{Coronet} cluster in CrA. With observations of the LkH$\alpha$101 cluster, \citet{ostenwolk2009} for the first time used simultaneous X-ray and multi-frequency radio observations to show that some sources show an inverse correlation between radio flux and spectral index. Most recently, observations of IC~348 and NGC~1333 have been carried out, but only few YSOS are detected in both bands (Forbrich, Osten, \& Wolk, \textit{submitted}). 

All these simultaneous measurements agree on the location of YSOs on an $L_X$ vs. $L_R$ plot. Both luminosities are among the highest observed toward stars. YSOs do not seem to fall directly on the GB relation, but they are shifted toward higher radio luminosities for a given X-ray luminosity. Still, there are only a few YSOs that seem to be clearly off the GB relation and their numbers are too low to provide firm conclusions. To date observations have been limited by the radio sensitivity, but the Expanded Very Large Array will soon lead to far more radio detections among YSOs. Higher signal-to-noise ratios will also allow us to better distinguish non-thermal (e.g., gyrosynchrotron) and thermal radio sources.
Interesting insights are also coming from VLBI radio observations (e.g., \citealp{dzib10}).

\vspace*{-6mm}
\section{The X-ray-Radio Disconnect in Ultracool Dwarfs} 

\noindent Prior to the first detection of radio emission from an
ultracool dwarf \citep{berger_2001}, measuring radio emission from
very low mass stars and brown dwarfs seemed improbable. Given the
tight correlation between radio and X-ray emission in higher-mass,
coronally-active stars, initial X-ray results predicted radio emission well below
detectable limits.  The detection of a radio flare from LP944-20
required a severe violation of the GB relation, with
$L_{\rm\nu,rad}/L_X \gtrapprox 10^{-11.5}$ Hz$^{-1}$.  Subsequent
radio detections of ultracool dwarfs \citep{berger_2002} have yielded
similar results, indicating a general violation of the correlation in
this regime.

With a large sample of
simultaneous observations, \citet{berger_2010} conclusively
demonstrated this correlation no longer holds for objects beyond
spectral type M6.  For M dwarfs in the range M0-M6, the
ratio of radio to X-ray emission is $L_{\rm\nu,rad}/L_X\approx
10^{-15.5}$.  This ratio steadily increases for cooler objects.  In
the range M7-M8 it is around $10^{-14}$ and beyond M9 it is greater
than $10^{-12}$.  It should be noted that there are a few objects
which have been detected in the X-rays without corresponding radio
emission, and hence may not violate the correlation.  However, with
the exception of a marginal detection from the L dwarf Kelu-1, these
are all earlier objects in the range M7-M9 and most do not have deep
radio luminosity limits and may yet be detected.

To investigate the nature of the breakdown of X-ray/radio correlation,
it useful to examine the trends of the X-ray and radio emission in
ultracool dwarfs separately. \citet{berger_2010} showed a sharp decline 
in X-ray activity for objects with spectral types beyond M7.  In contrast,
they also found indications for an increase in the ratio activity ($L_{\rm rad}/L_{\rm
bol}$) for those objects cooler than M7.  It is clear, then, that the
breakdown is the due to the decline in X-ray luminosity combined with
the sustained strength of the radio emission.  But why do particle
acceleration and plasma heating no longer correlate?  As demonstrated
by the radio emission, neither magnetic field dissipation nor particle
acceleration appear to be affected by the increasingly neutral
atmospheres of ultracool dwarfs.  The total radio luminosities remain
roughly constant even in these cooler objects, indicating the fraction
of magnetic energy that goes into accelerating the electrons
responsible for the radio-emission does not change, nor does the
efficiency of field dissipation.

The difference appears to lie in the efficiency of the plasma heating,
which is responsible for the hot X-ray producing gas.  If the
radio-emitting electrons are directly responsible for plasma heating
in higher mass stars, the enhanced trapping of these electrons could
account for the breakdown in the correlation.  It could also be
produced by a change in the geometry of the radio-emitting regions,
should they evolve to smaller sizes in lower-mass objects and hence
have less of an effect on the large-scale coronal heating.  However,
this is unlikely given that rotationally stable, quiescent radio
emission and periodic H$\alpha$ emission has been detected from
several ultracool dwarfs, indicative of large magnetic filling factors.

Since radio emission requires only a small population of relativistic
electrons, {\it a decline in the bulk coronal density could suppress the
X-ray heating without affecting the radio emission}.  The decreased
X-ray emission, if caused by a loss in coronal density, could be
attributed to coronal stripping.  There are hints of super-saturation
in the X-ray-rotation relation in the fastest rotators among the
ultracool dwarfs \citep{berger_2008}. For objects later than M7, there is decline in $L_X/L_{\rm bol}$ for objects with decreasing rotation periods.  The
median value of $L_X/L_{\rm bol}$ is $\approx 10^{-4}$ for $P > 0.3$
d, while for $P < 0.3$ d it is $L_X/L_{\rm bol}\approx 10^{-5}$. The
combination of rapid rotation and the shrinking co-rotation radius in
lower mass stars may lead to the decrease in X-ray emission among
these objects.  In comparision, there are indications that the fastest
rotators among ultracool dwarfs are more likely to be detected in the
radio and are the most severe violaters of the X-ray-radio correlation
\citep{berger_2008}.

In contrast to hotter stars which primarily produce radio emission
through gyrosynchrotron radiation, several
ultracool dwarfs have been observed with short-duration, highly
polarized, narrow band, bursts which can be attributed to a coherent
emission process, such as the electron cyclotron maser (ECM)
instability \citep{benz01,hallinan_2007,hallinan_2008,berger_2009}.  Although
this process cannot account for all radio emission detected from
ultracool dwarfs, this change in emission mechanism may be indicative
of a change in the properties of the relativistic electron population
and its impact on large-scale coronal heating.

\vspace*{-5mm}
\section{Conclusions}

In short, we are not headed toward a divorce. Instead, the X-ray and radio luminosities of cool stars
appear to be in an ``open relationship'' where a lot depends on the type of radio emission that is present.
Until now, observational data have not always been sensitive enough to unambiguously identify dominant emission mechanisms, for example in the case of YSOs.
Clearly, the $L_{\rm X}/L_{\rm R}$ relation does not apply to all stars. As mentioned above, inactive stars violate this relation as they do not show non-thermal radio (gyrosynchrotron) emission. The ``non-flaring'' Sun is an example. 
Either, there are additional heating mechanisms at work in these stars that 
do not involve high-energy electrons, or the energy transformation process 
is more efficient in heating the plasma to sufficiently high temperatures so
as to become visible in X-rays.
The other important class of stars violating the relation are brown dwarfs, but coherent radio radiation mechanisms may matter here.
Similarly, a number of protostellar objects do not follow the standard trend although they are magnetospheric radio and X-ray sources. Here, however, the measurement of either of the luminosities is difficult. Radio gyrosynchrotron emission may be attenuated by overlying ionized winds (e.g., the jets easily detected as thermal radio sources); X-ray emission could be partially attenuated by neutral gas masses, such as neutral winds, accretion streams, or molecular outflows. If only part of the coronal emission is (fully) attenuated, e.g., by accretion streams, then an assessment of the intrinsic luminosities becomes impossible.
Further observations, particularly deeper radio observations with new
instruments such as EVLA, as well as large X-ray surveys with
\emph{Chandra} and XMM-\emph{Newton} will help shed light on the
limits of the radio/X-ray correlation in cool stars.

\vspace*{-6mm}
\section*{Appendix: What is the connection between nonthermal radio emission and thermal 
X-ray emission?}

The lack of consensus among the theoreticians concerning which physical models
best explain heating and particle acceleration
and the GB correlation 
suggest to Jeff Linsky that there must be a simple approach
to understanding the relevant physics. Given that
flares occur when there are rapid changes in the
structure of complex magnetic fields, Maxwell's equations require the presence
of electric fields and currents. In a plasma the acceleration force of the
electric field $F_{accel}=eE$ is opposed by a collisional drag force
$F_{drag}\sim (n_e/T)(v_{th}/v)^2$ when the speed $v$ exceeds the thermal speed
$v_{th}=(kT/m)^{1/2}$. Those electrons travelling at greater than a critical
speed $v_c\sim (n_e/E)^{1/2}$ set by $F_{accel}>F_{drag}$, will be accelerated 
to
high speeds, a process called ''runaway''. If the electric field exceeds the
Dreicer field, $E>E_D=4\pi e^3 ln\Lambda/kT$ \citep{Holman1985}, 
then all electrons will  
runaway, but this rarely occurs because large E 
fields produce turbulence and anomalous resistivity \citep{Norman1978}.

When $E<E_D$ only those electrons in the tail of the Maxwell-Boltzmann 
velocity distribution, $f(v)$, with $v>v_c$
are accelerated. In this ``sub-Dreicer regime'', the ratio of nonthermal
to thermal electrons is 
$N_{nonth}/N_{th}=\int_{v_c}^{\infty}f(v)dv/\int_0^{\infty}f(v)dv$. When
$E\ll E_D$, $N_{nonth}/N_{th}\approx 0.5e^{-0.5[(E_D/E)^{1/2}-E/E_D]^2}$
\citep{Norman1978}. In addition to 
accelerating electrons, electric fields can heat the plasma by several 
mechanisms, the simplest being the Joule heating rate $iE=i^2r$, where $i$ is 
the current and the resistance $r$ can be anomalous due to current-driven 
instabilities. \citet{Holman1985} developed the theory of heating and 
acceleration by electric fields in the context of solar flares. For the
example of $T=10^7$~K, $B=300$~G, and $EM=10^{45}$~cm$^{-3}$, 
he finds that the time 
scales for Joule heating and acceleration of $10^{32}$ electrons are both 
about 30 seconds and the energy that goes into particle acceleration is 
always less that the heating rate.
In his Case I, the assumption of $E=0.03E_D$ corresponds to 
$v_c=5.8v_{th}$, electron acceleration to maximum energy $W_{max}=10$~MeV, and
$N_{nonth}/N_{th}=3.4\times 10^{-8}$. 
In his Case II, doubling the electric field 
strength to $E=0.06E_D$ results in $v_c=4.2v_{th}$, $W_{max}=100$~keV, and
$N_{nonth}/N_{th}=1.5\times 10^{-4}$. Thus the $N_{nonth}/N_{th}$ ratio and 
the maximum energy of the nonthermal electrons both depend very 
strongly on $E/E_D$, but in the opposite sense.

Since gyrosynchrotron radiation is broad band with 
$\Delta \nu\approx 1\times 10^{10}$ Hz, the ratio of radio to X-ray
emission ratio is $R=L_R\Delta\nu/L_X = 10^{-5.5\pm 0.5}$, much less than 
unity as predicted by \citet{Holman1985}. Since the gyrosynchrotron radio 
emission rate per relativistic electron is proportional to its energy squared,
the main contribution to $L_R\Delta\nu$ will be from the highest energy 
electrons. Free-free X-ray emission from $T=10^7$~K electrons is proportional 
to $N_{th}^2V$. Thus $R$ is proportional to 
$(N_{nonth}/N_{th})W_{max}^2V_J/N_{th}V$, 
where $V$ is the volume of the thermal gas from which the nonthermal 
electrons have been swept up and $V_J$ is the volume of the nonthermal 
electrons in the current channel.
For Holman's Case I, $V_J/V=10^{-4}$ and  
$R\propto3.4\times10^{-8}(10^7{\rm eV})^2 10^{-4}/10^9{\rm cm}^{-3}=
3.4\times10^{-7}$.
For Case II, $V_J/V=1/200$ and  
$R\propto1.5\times10^{-4}(10^5{\rm eV})^2(1/200)/10^{11}{\rm cm}^{-3}=
0.75\times10^{-7}$. 
The similar ratios for the two theoretical cases, despite the very different 
$N_{nonth}/N_{th}$ values, show that DC electric fields 
can produce both runaway electrons in the sub-Dreicer
regime and Joule heating consistent with a constant $L_R/L_X$ relation.



\end{document}